\documentclass[preprint2]{aastex}

\slugcomment{}

\shorttitle{X-ray Emission from G32.8$-$0.1}
\shortauthors{Bamba et al.}

\begin{document}

\title{Discovery of X-ray Emission
from the Galactic Supernova Remnant G32.8$-$0.1
with {\it Suzaku}}

\author{Aya Bamba\altaffilmark{1},
Yukikatsu Terada\altaffilmark{2},
John Hewitt\altaffilmark{3},
Robert Petre\altaffilmark{3},
Lorella Angelini\altaffilmark{3},
Samar Safi-Harb\altaffilmark{4},
Ping Zhou\altaffilmark{5},
Fabrizio Bocchino\altaffilmark{6},
Makoto Sawada\altaffilmark{1}
}
\altaffiltext{1}{
Department of Physics and Mathematics, Aoyama Gakuin University
5-10-1 Fuchinobe Chuo-ku, Sagamihara,
Kanagawa 252-5258, Japan
}
\altaffiltext{2}{
Department of Physics, Science, Saitama University, Sakura, Saitama 338-8570,
Japan
}
\altaffiltext{3}{
NASA Goddard Space Flight Center, Greenbelt, MD 20771, USA
}
\altaffiltext{4}{
Department of Physics and Astronomy, University of Manitoba,
Winnipeg MB R3T 2N2, Canada
}
\altaffiltext{5}{
Department of Astronomy, Nanjing University, Nanjing 210093, China
}
\altaffiltext{6}{
INAF---Osservatorio Astronomico di Palermo, Piazza del Parlamento 1,
90134, Palermo, Italy
}

\begin{abstract}
We present the first dedicated X-ray study of the supernova remnant (SNR)
%A detailed analysis of the X-ray emission from 
%the supernova remnant (SNR) 
G32.8$-$0.1 (Kes~78) with {\it Suzaku}.
% is presented.
X-ray emission from the whole SNR shell 
has been detected for the first time.
The X-ray morphology is well correlated with the emission from the radio shell, while anti-correlated with
the molecular cloud found in the SNR field.
The X-ray spectrum shows not only conventional low-temperature 
%thermal emission
($kT \sim 0.6$~keV) thermal emission in a non-equilibrium ionization state,
but also a very high temperature
% ($kT \sim 3$~keV) 
 ($kT \sim 3.4$~keV)
component with a very low ionization timescale ($\sim 2.7 \times10^9$ cm$^{-3}$~s), or a hard non-thermal component
with a photon index $\Gamma$$\sim$2.3.
The average density of the low-temperature plasma is rather low,
of the order of $10^{-3}$--$10^{-2}$~cm$^{-3}$,
implying that this SNR is expanding into a low-density cavity.
We discuss the X-ray emission of the SNR, also detected in TeV with H.E.S.S., 
together with multi-wavelength studies of the remnant and
other gamma-ray emitting SNRs, such as W28 and RCW~86.
%Together with these results,
%we consider that G32.8$-$0.1 is similar source to
%the gamma-ray SNRs such as W28 or RCW~86.
%
Analysis of a time-variable source,
2XMM~J185114.3$-$000004, found in the northern part of the SNR, 
is also reported for the first time.
Rapid time variability and a heavily absorbed hard X-ray spectrum
suggest that this source could be a new supergiant fast X-ray transient.
\end{abstract}

\keywords{%acceleration of particles
ISM: supernova remnants
--- ISM: individual (\objectname{G32.8$-$0.1, Kes~78})
--- X-rays: ISM
--- stars: neutron
--- stars: individual (\objectname{2XMM~J185114.3$-$000004})
}

\section{Introduction}

Supernova remnants (SNRs) are believed to be the primary sites for cosmic ray acceleration up to the `knee' of the cosmic rays spectrum.
%one of the most prausible site of 
%cosmic-ray accelelation.
X-ray observations revealed that
shells of several young SNRs are synchrotron X-ray emitters,
implying that they are the acceleration sites of particles
\citep{koyama1995,koyama1997}.
On the other hand,
the number of SNRs with a synchrotron X-ray emitting shell is limited
\citep{nakamura2012}.
Recent very high energy (VHE) gamma-ray observations
with {\it H.E.S.S.}, {\it MAGIC}, and {\it VERITAS} are continually revealing
SNRs\footnote{see http://www.physics.umanitoba.ca/snr/SNRcat for a compilation of
X-ray and gamma-ray observations of all Galactic SNRs.} as sites for
% that
%there are
energetic particles accelerated at SNR shocks up to the TeV range
% on shocks of young SNRs.
\citep{aharonian2004,aharonian2007,aharonian2009,albert2007,acciari2010}.
Furthermore, recent {\it Fermi} observations show that,
not only young, but also middle-aged SNRs are
GeV gamma-ray emitters
\citep{abdo2010w44,abdo2010ic443,abdo2009w51c,funk2011}.
%An interesting fact is that
%many GeV emitters detected with {\it Fermi}  are
%categorized as mixed-morphology (MM) type SNRs
%\citep{yusef-zadeh2003,hewitt2009};
%i.e. SNRs with a shell-like structure in the radio, but with a centrally-filled morphology in thermal X-rays
%\citep{rho1998}\footnote{see http://www.physics.umanitoba.ca/snr/SNRcat for the high-energy catalogue of SNRs
%compiling X-ray and $\gamma$-ray observations, including the mixed-morphology class.}.
%note: FYI Green calls them still shell-type SNRs..
%Moreover, the X-ray emitting plasma in many MM SNRs shows over-ionized plasma
%\citep[][for example]{uchida2015},
%although there are some exceptions such as W28 and RCW~86
%which has a shell-like morphology not only in radio
%but also in X-rays \citep{nakamura2014,bamba2000}.
%In order to understand what causes such differences,
Some of these gamma-ray emitting SNRs are not covered
by deep X-ray observations.
We need a larger sample of X-ray studied SNRs with GeV and VHE gamma-ray
emission
to understand the nature of these cosmic ray accelerators.

G32.8$-$0.1 (Kes~78) was discovered by \citet{velusamy1974}
in the radio band at 11~cm wavelength.
OH masers were detected from the SNR \citep{green1997},
suggesting an interaction with an adjacent molecular cloud
\citep{koralesky1998}.
Observations of $^{12}$CO \citep{zhou2007,zhou2011}
reveal a dense molecular cloud
on the eastern side of the SNR.
%Molecular cloud on the east of the SNR is observed by
%\citet{zhou2007,zhou2011}
%and 
\citet{zhou2011} derived a kinematic distance to the SNR
of 
%showed the distance to the SNR is 
4.8~kpc.
Significant GeV emission was also found close to this SNR, with 
2FGL~J1850.7$-$0014 in the 2nd Fermi source Catalog \citep{nolan2012}
suggested to be related to G32.8$-$0.1.
More recently, \citet{auchettl2014}
studied G32.8$-$0.1 using 52 months of data with {\it Fermi};
however, given the uncertainties in the $\gamma$-ray background model and contamination
by other nearby sources, they were unable to confirm the excess of GeV emission from the SNR.
The 3rd Fermi source catalog \citep{acero2015} confirmed the source again
and revised the position and its error.
A VHE extended gamma-ray source,
HESS~J1852$-$000, was found by the H.E.S.S. team
outside the eastern edge of the remnant \citep{kosack2011}\footnote{http://www.mpi-hd.mpg.de/hfm/HESS/pages/home/som/2011/02/}.
This emission partly overlaps with the radio shell of the SNR and with the molecular cloud seen in CO.
While the interaction between the SNR and the molecular cloud had been suggested as a plausible scenario
for the TeV emission seen with H.E.S.S., an alternative, but less likely, scenario proposed was its association
with a pulsar wind nebula (PWN) associated with a nearby pulsar (PSR~J1853--0004).
The gamma-ray emission from the SNR implies that
there is some high-energy activity from this remnant, despite its nature being still unresolved.
%on this remnant
This SNR therefore provides another example SNR with potential GeV and with VHE gamma-ray emission.
In X-rays, the only information we have so far published on the remnant comes from an  {\it XMM-Newton} study
of the northern part of the SNR shell \citep{zhou2011}.
% has been detected significant thermal X-rays
%from the northern part of the shell with {\it XMM-Newton}.
We still lack an X-ray detection of the whole remnant
which is necessary to understand the properties of this SNR and shed light
on its multi-wavelength emission.

In this paper, we report on the first detailed X-ray imaging and spectroscopy study
of the entire SNR, G32.8$-$0.1,
using {\it Suzaku} \citep{mitsuda2007}.
We also report on a transient source which went into outburst during our observation.
The observation details are summarized in \S\ref{sec:obs}.
A first analysis of the {\it Suzaku} X-ray data for these sources is
presented in \S\ref{sec:results},
the results of which are discussed in \S\ref{sec:discuss}.

\section{Observations and Data Reduction}
\label{sec:obs}

G32.8$-$0.1 was observed by {\it Suzaku} with two pointings,
on 2011, Apr. 20--22.
The coordinates of two pointings are listed in Table~\ref{tab:obslog}.
{\it Suzaku} has two active instruments:
four X-ray Imaging Spectrometers (XIS0--XIS3; \cite{koyama2007}), with
each at the focus of an X-Ray Telescope (XRT; \cite{serlemitsos2007}),
and a separate Hard X-ray Detector (HXD; \cite{takahashi2007}).
Only three XISs could be operated for this study due to a problem with XIS2.
XIS1 is a back-illuminated CCD,
whereas the others are front-illuminated.
The XIS instruments were operated in normal full-frame clocking mode
with spaced-row charge injection \citep{nakajima2008,uchiyama2009},
whereas the HXD was operated in normal mode. 
Data reduction and analysis were made
with HEADAS software version 6.13
and XSPEC version 12.8.0.
The data was reprocessed with the
calibration database version 2013-03-05 for XIS,
2011-09-05 for HXD,
and 2011-06-30 for XRT.

In the XIS data screening,
we followed the standard screening criteria;
filtering out data acquired during passage
through the South Atlantic Anomaly (SAA),
with an elevation angle to the Earth's dark limb below 5~deg,
or with elevation angle to the bright limb below 25~deg
in order to avoid contamination by emission from the bright limb.
Table~\ref{tab:obslog} shows the remaining exposure time.

As for the HXD dataset,
we also followed ths standard screening criteria;
filtering out data obtained during passage through the SAA,
with an elevation angle to the Earth's limb below 5~deg,
and cutoff rigidity smaller than 8~GV.
The resultant exposure time for each observation is shown
in Table~\ref{tab:obslog}.
We adopted the LCFIT model of \citet{fukazawa2009}
for the non-X-ray background (NXB) model.
The cosmic X-ray background (CXB) flux is estimated from
the {\it HEAO1} results \citep{boldt1987},
and treated as an additional background component.

\section{Results}
\label{sec:results}

\subsection{Images}

\begin{figure}
\epsscale{0.8}
\plotone{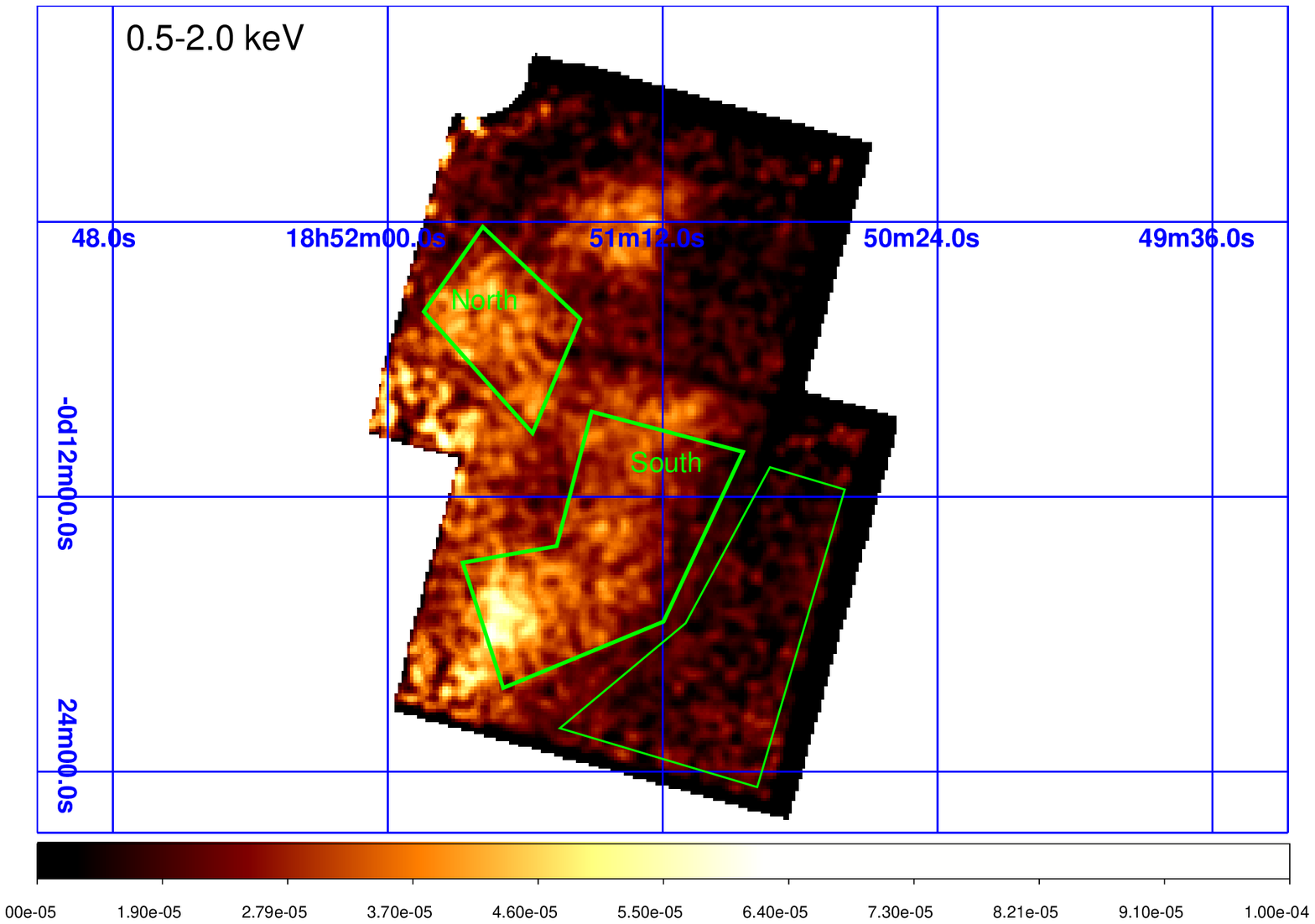}
\plotone{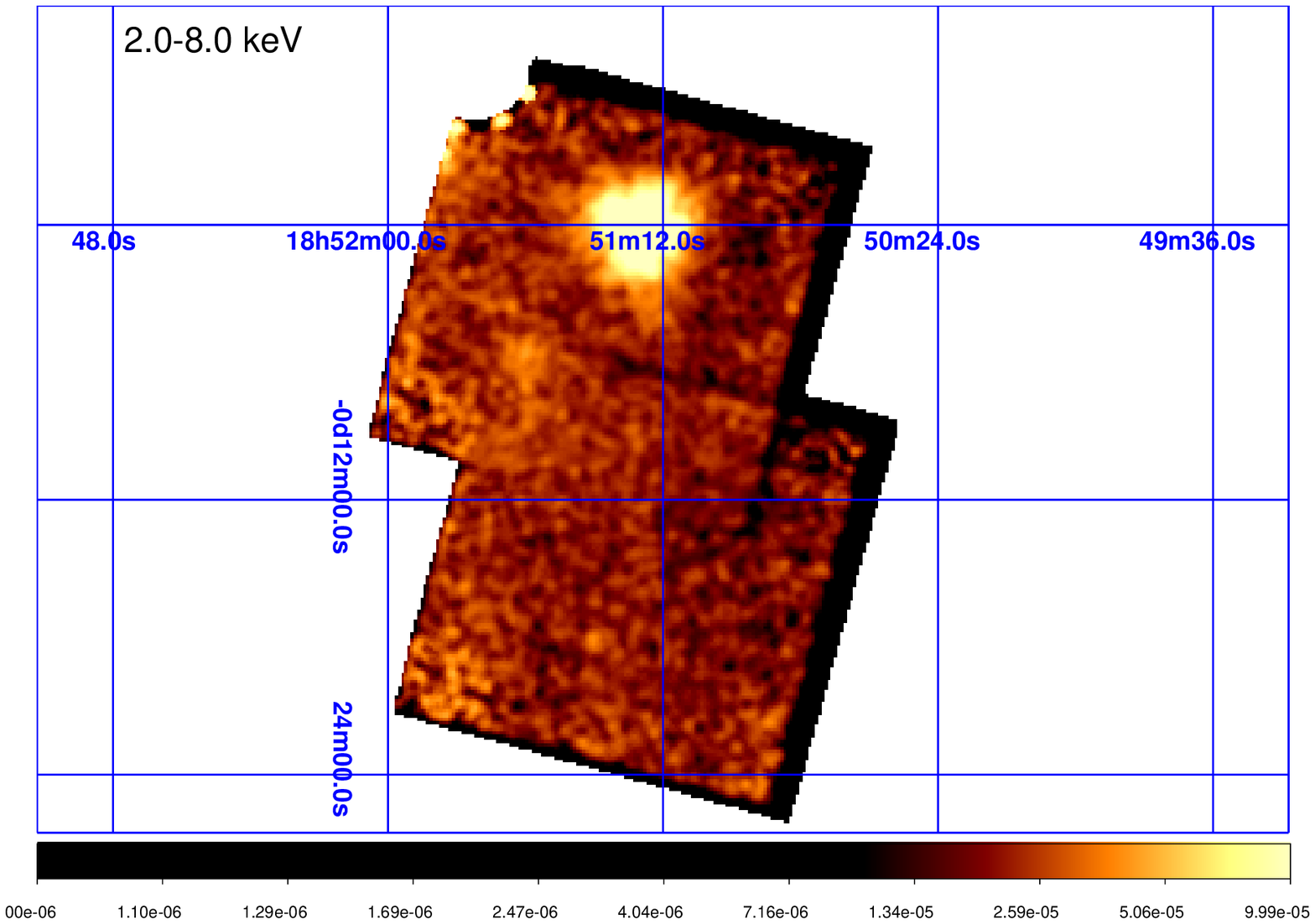}
\caption{The 0.5--2.0~keV (top) and 2.0--8.0~keV (bottom) band images of
the SNR G32.8$-$0.1.
These images were smoothed using a Gaussian filter of
Kernel=0.4~arcmin.
Thick and thin green regions represent
the source and background regions for the spectral analysis, respectively.
The bright point source clearly visible in the hard band on the northwest of the remnant 
is 2XMM~J185114.3$-$000004.
}
\label{fig:xrayimage}
\end{figure}

The XIS 0.5--2.0~keV and 2.0--8.0~keV mosaic images
are shown in figure~\ref{fig:xrayimage}
The vignetting has been corrected in each image
using {\it xissim} \citep{ishisaki2007} after subtracting the NXB
% non X-ray background (NXB) 
\citep{tawa2008}.
One can see clearly a clumpy shell-like structure
elongated in the north-south direction
in the 0.5--2.0~keV band image.
On the other hand, the 2--8~keV band image is dominated by
a bright point source detected in our observation in the northern part of the remnant.
We find that this point source is positionally coincident with
the second {\it XMM-Newton} serendipitous source catalog source,
2XMM~J185114.3$-$000004 \citep{watson2009}.
This source is now cataloged in the 3XMM Data Release 5
(http://xmm-catalog.irap.omp.eu; Zolotukhin et al., in prep.)
as 3XMM~J1851$-$000002,
with the best position of (Ra, Dec.) = (282.80961, $-$0.00080)
with the position error od 0.63~arcsec.
In this paper,
we stick to the 2XMM name
since the SIMBAD database uses only 2XMM source lists.
%In order to examine the hard X-ray diffuse emission,
%we made 2.0--8.0~keV band image
%without the point source region
%as shown in Figure~\ref{fig:xrayimage}.
%One can see that the interior of the SNR is brighter than the outside
%in the hard X-ray band,
%although the statistics is very limited.

\begin{figure}

\epsscale{0.5}
\plotone{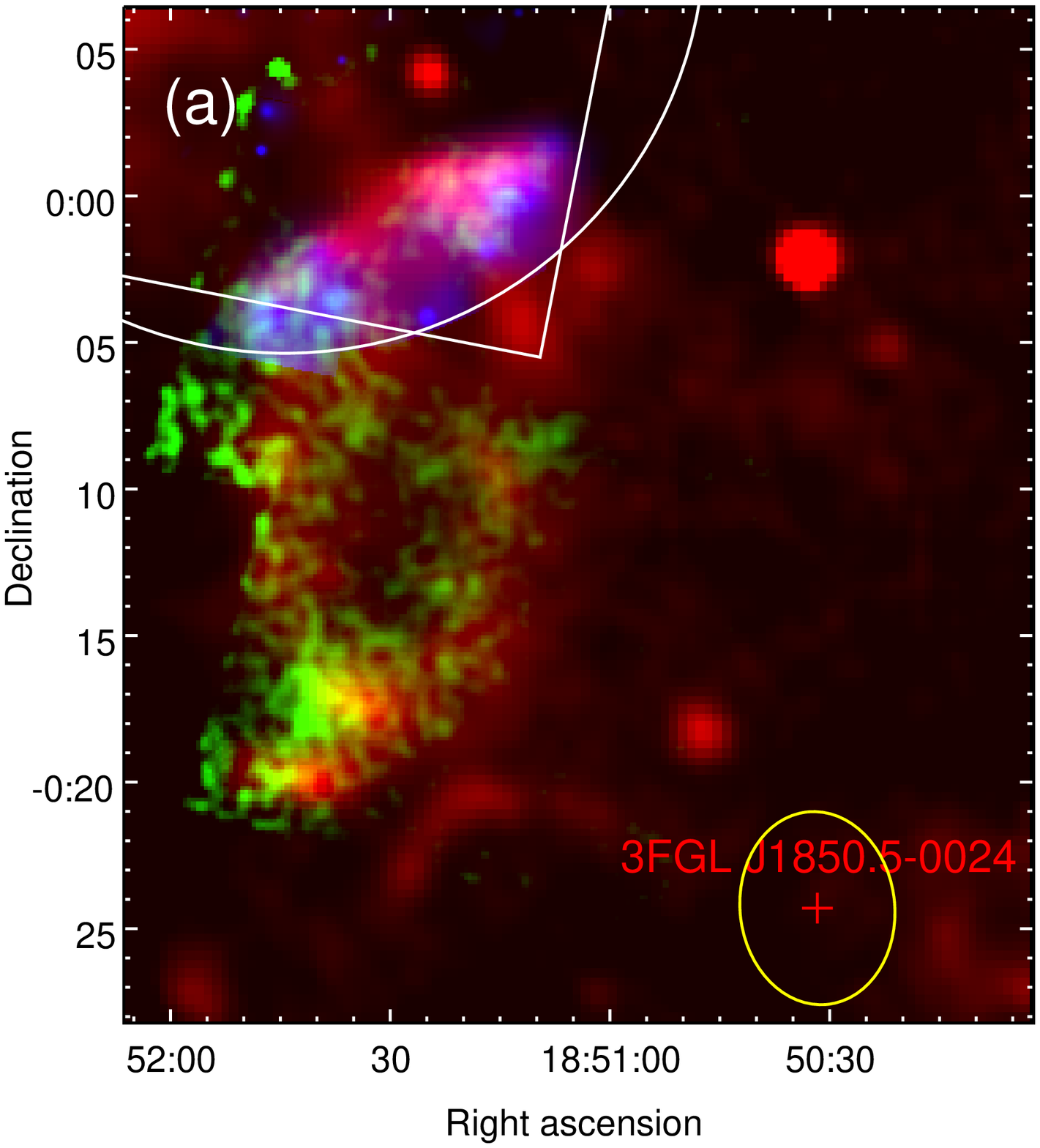}
\plotone{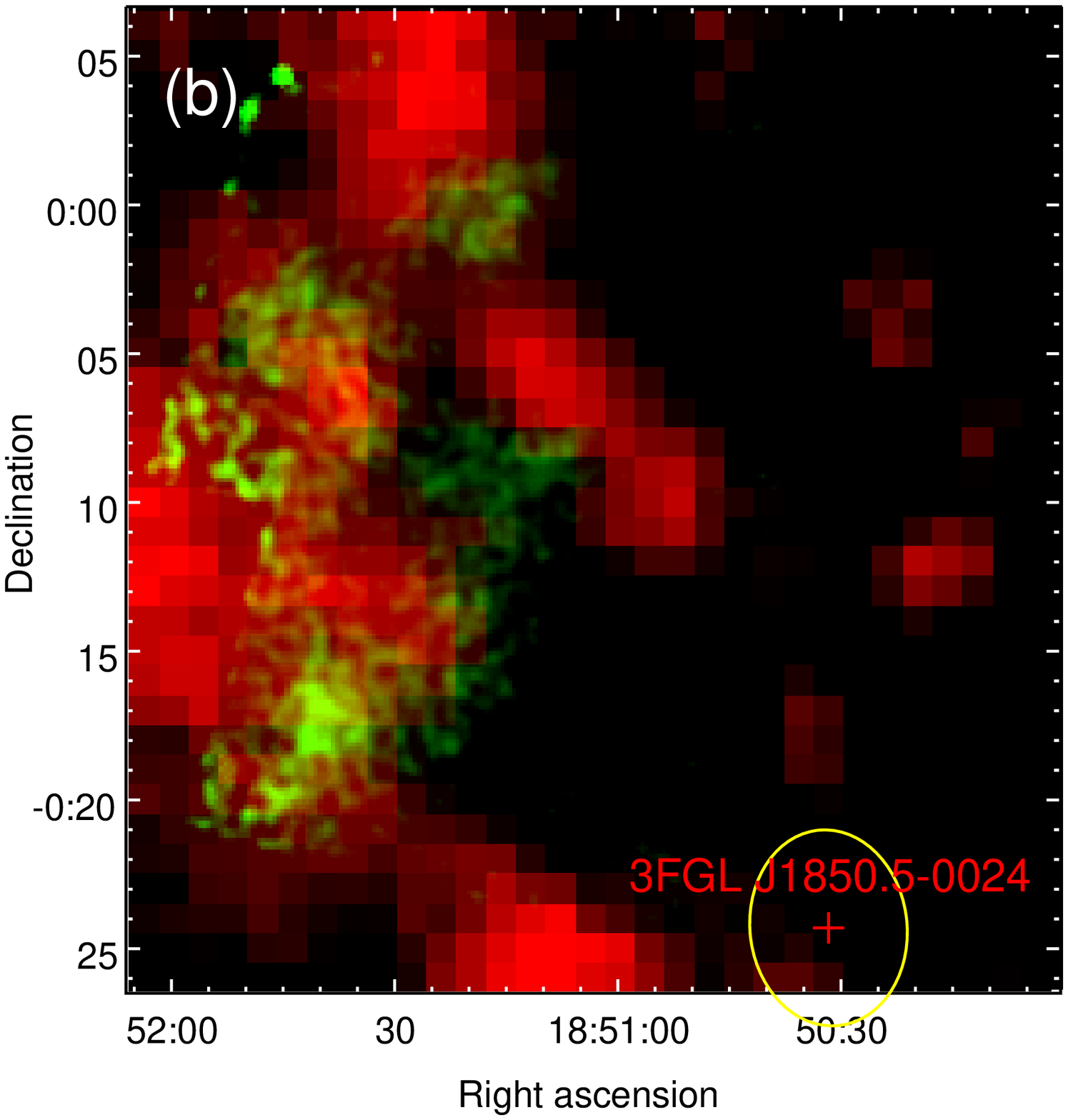}
\caption{%
(a) VGPS 1.4 GHz continuum \citep{stil2006} (red)
together with the 0.5--2~keV {\it Suzaku} (green),
the 0.5--8~keV {\it XMM} (ObsID=0017740501; \cite{zhou2011}) (blue),
and the 3FGL source region \citep{acero2015} (yellow).
White circle and box represent the {\it XMM-Newton} MOS and pn field of views,
respectively.
(b) PMOD $^{12}$CO~(J=1--0) image in the velocity range of 80--84~km~s$^{-1}$
\citep{zhou2011} (red)
with the 0.5--2~keV {\it Suzaku} (green)
and the {\it Fermi} source region \citep{acero2015} (yellow).
}
\label{fig:multi-band}
\end{figure}

Figure~\ref{fig:multi-band}(a) shows
archival VLA Galactic Plane Survey (VGPS) continuum data at 1.4~GHz \citep{stil2006}
together with the 0.5--2~keV {\it Suzaku}, 0.5--8~keV {\it XMM-Newton} \citep{zhou2011},
and the {\it Fermi} source region.
We highlight in this image
%revealed that the
the diffuse X-ray emission detected by {\it XMM-Newton}
is the northern part of the SNR shell.
As seen in Figure~\ref{fig:multi-band}(a), the X-ray emission traces
the radio shell emission; in particular the {\it Suzaku} emission follows
the morphology of the elongated bright radio emission in the southern part of the remnant.
%well trace the radio shell,
%together with coincidence of the southern bright part in both images.

Figure~\ref{fig:multi-band}(b) shows the
%on the other hand,
$^{12}$CO (J=1--0) image in the velocity range of
80--84~km~s$^{-1}$
taken by the Purple Mountain Observatory at Delingha (PMOD) in China
\citep{zhou2011},
with the 0.5--2~keV {\it Suzaku} image and the {\it Fermi} source region overlapped.
This image reveals the molecular cloud emission surrounding the X-ray shell.
We note that the dent seen in the eastern part of the shell is likely caused
by the interaction between the SNR and the molecular cloud.
% eastern X-ray shell hit the molecular cloud enhancement
%which causes the dent of the shell in this region.
On the other hand, the elongation towards the south may be caused by the expansion
of the shell into a relatively lower density medium void of the molecular cloud.
%south-western shell expands free from the interaction,
%which may cause the larger radius in this direction.
The position of the GeV emission is on the western part,
and although the position error is large,
%there is no clear association between the GeV source and the molecular cloud plus X-ray emission to its east
there is no overlap
between the GeV source and the molecular cloud plus X-ray emission to its east
\citep{acero2015}.

\subsection{Spectra of diffuse emission}

Here, we present our
%have performed
spatially resolved spectroscopic study of the diffuse emission
below 10~keV with XIS.
We divided the remnant into two parts (north and south) and selected a nearby
background region, as shown in Figure~\ref{fig:xrayimage}.

\subsubsection{Background spectrum}

The simplest way for the background estimation in the source regions is
%just 
to subtract the spectrum extracted from the background region,
but this sometimes introduces an uncertainty due to vignetting.
We thus reproduce the background spectrum in the source region
as described next.

The background emission contains the non-X-ray background (NXB), plus the 
cosmic X-ray background (CXB), galactic ridge X-ray emission (GRXE),
and emission from the local hot bubble (LHB).
The first component is uniform in the field of view,
whereas the others are affected by vignetting.
The NXB in the background region is
generated by {\tt xisnxbgen} \citep{tawa2008}, and 
was subtracted from the spectrum
after adjusting the normalization above 10~keV,
where we expect no emission except for the NXB.

\begin{figure}
\epsscale{0.5}
\plotone{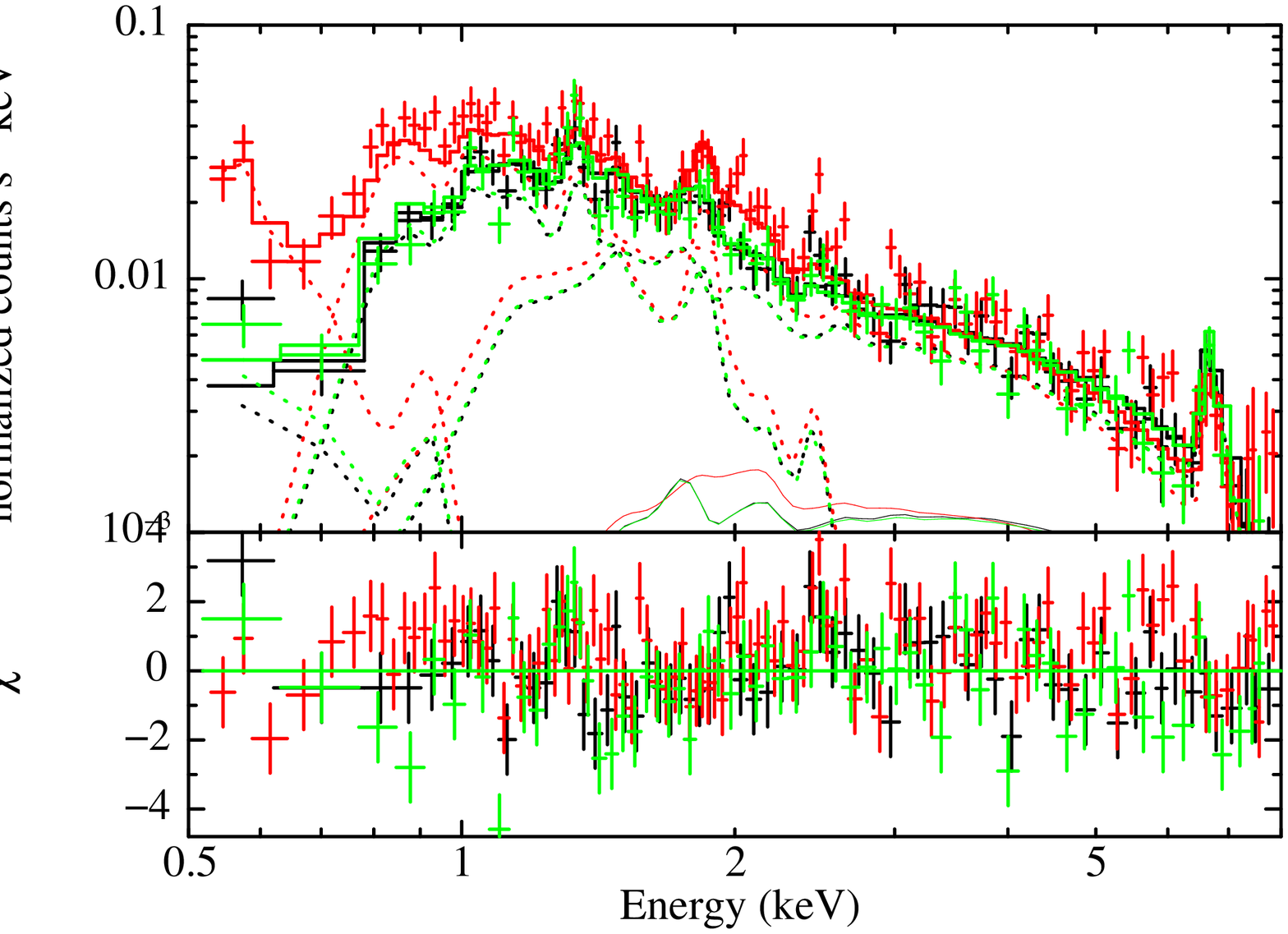}
\caption{%
NXB-subtracted spectra of the background region.
Black, red, and green crosses represent XIS0, 1, and 3 spectra,
respectively.
Dotted and thin solid lines represent the
thermal and CXB components, respectively.}
\label{fig:background-spec}
\end{figure}

The NXB-subtracted spectra is shown in Figure~\ref{fig:background-spec}.
We fit it with an absorbed CXB + GRXE + LHB model.
To reproduce the CXB emission,
we assumed the power-law shape with a photon index of 1.4,
and the surface brightness in the 2--10~keV band of 
5.4$\times 10^{-15}$~erg~s$^{-1}$cm$^{-2}$arcmin$^{-2}$
\citep{ueda1999}.
Note that the background region has an area of 73.7~arcmin$^2$.
The hydrogen column density of the CXB in this direction was fixed to
$1.82\times10^{22}$~cm$^{-2}$, as
determined by H${\rm I}$ observations \citep{dickey1990}.
We also added an absorbed thermal $apec$ component for 
the emission from the LHB \citep{yoshino2009},
and an absorbed two-temperature $apec$ component
for the GRXE emission \citep[for example, ][]{koyama2007b}.
The best-fit model and parameters are shown in 
Figure~\ref{fig:background-spec} and Table~\ref{tab:background-spec},
respectively.
The best-fit parameters are roughly consistent with
the known components in other regions
\citep{yoshino2009,ryu2009,sawada2011}.

Assuming that the background spectrum has the same shape and surface-brightness
in the background and nearby source regions,
we simulate the background spectrum in
addition to the NXB spectrum in the source region,
using the {\tt fakeit} command in XSPEC.
We assumed a similar exposure for the simulated background as for
the source spectra.
This is a similar method to what's described in \citet{fujinaga2013}.
We then use this background spectrum in the following analysis.

\subsubsection{Source spectra}

We initially perform the spectral analysis of the global SNR
using the total emission detected with $\it{Suzaku}$
from the SNR (i.e. the North+South regions shown in Fig.~1).

\begin{figure}
\epsscale{0.5}
\plotone{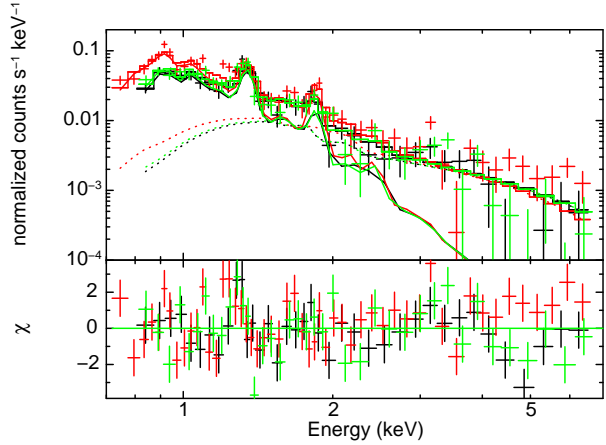}
\caption{%
Background-subtracted spectra of the total diffuse emission.
Black, red, and green crosses represent XIS0, 1, and 3 spectra,
respectively.
Solid and dotted lines represent the {\tt vnei} and power-law component,
respectively.
}
\label{fig:diffuse_spectra}
\end{figure}

Figure~\ref{fig:diffuse_spectra} shows the background-subtracted spectra
for the total emission.
The spectrum shows emission lines from highly ionized Ne, Mg, and Si,
implying that the emission contains a thermal component.
We thus fit the spectra initially with an absorbed thermal model
from collisionally-ionized plasma ({\tt apec} in XSPEC).
For the absorption model, we applied the {\sc phabs} model,
%in this paper.
%The absorption model i
which includes the cross sections of
\citet{balcinska-church1992}
with solar abundances \citet{anders1989}.
The fit was rejected with a reduced $\chi^2$/d.o.f. of 
405.4/122 even when the abundances of metals were
 treated as free parameters.
Introducing a non-equilibrium ionization model
({\tt vnei} and {\tt vvrnei} in XSPEC),
we got a better reduced $\chi^2$/d.o.f. of 232.0/122 and
381.1/123, respectively, but
the fit was still rejected.
The best fit we found with the {\tt nei} model
required a very high temperature,
$kT$$\sim$3.4~keV, for the SNR as shown in Table~\ref{tab:total-spec}.
This result suggests that the emission contains a hard
tail.

We then tried a two-temperature {\tt nei} model and
a single temperature {\tt nei} model plus a power-law component.
The former model ({\tt nei}+{\tt nei}) returned a smaller reduced $\chi^2$ of
208.6/122,
but required a high temperature ($>$2~keV) and 
a small ionization timescale ($n_e t$ $\sim 10^9$~cm$^{-2}$~s).
On the other hand, the ({\tt nei} + power-law) component model,
while giving a slightly larger  $\chi^2$ of 223.1/121,
yielded parameters that are reasonable for X-ray emission from an SNR.
With the available data and complications from the background emission,
we can not easily conclude whether the hard tail is thermal or non-thermal,
and so we discuss both models in the following section.
The best-fit models and parameters are shown in Table~\ref{tab:total-spec}
and in Figure~\ref{fig:diffuse_spectra} (for the {\tt vnei}+power-law model).

We have made same analysis with another plasma model,
{\tt vpshock}, and found that
the results are basically same with the same best-fit parameters
and reduced $\chi^2$ within errors.

We further verified whether the hard-tail component is real or not.
The residuals in the hard X-ray band remain
even when we subtract the background photons
accumulated from the background region directly.
The background was simulated with the best-fit parameters
in the background region fitting (Table~\ref{tab:background-spec}),
which may introduce systematic uncertainty.
We tried fitting with another simulated background set
with different parameters within the range of Table~\ref{tab:background-spec},
and found that the results for the source spectra do not change within error.
Under-estimation of the NXB component can cause such a residual emission,
but the NXB count rate in the 3--5~keV
(mainly from the power-law component by CXB, Figure~\ref{fig:diffuse_spectra})
is around 10\% of that of the total emission,
thus it is difficult to consider the hard-tail component as due to
the mis-estimation of the NXB,
since the NXB reproduceability is around 10\% \citep{tawa2008}.
The possibility of the contamination from the bright and hard northern point source,
2XMM~J185114.3$-$000004, is unlikely
since the best-fit photon index of the diffuse emission is much softer than that of
the point source (see \S\ref{sec:2XMM}).
We thus conclude that the hard-tail component most likely originates from the 
SNR shell.

\begin{figure}
\epsscale{0.45}
\plotone{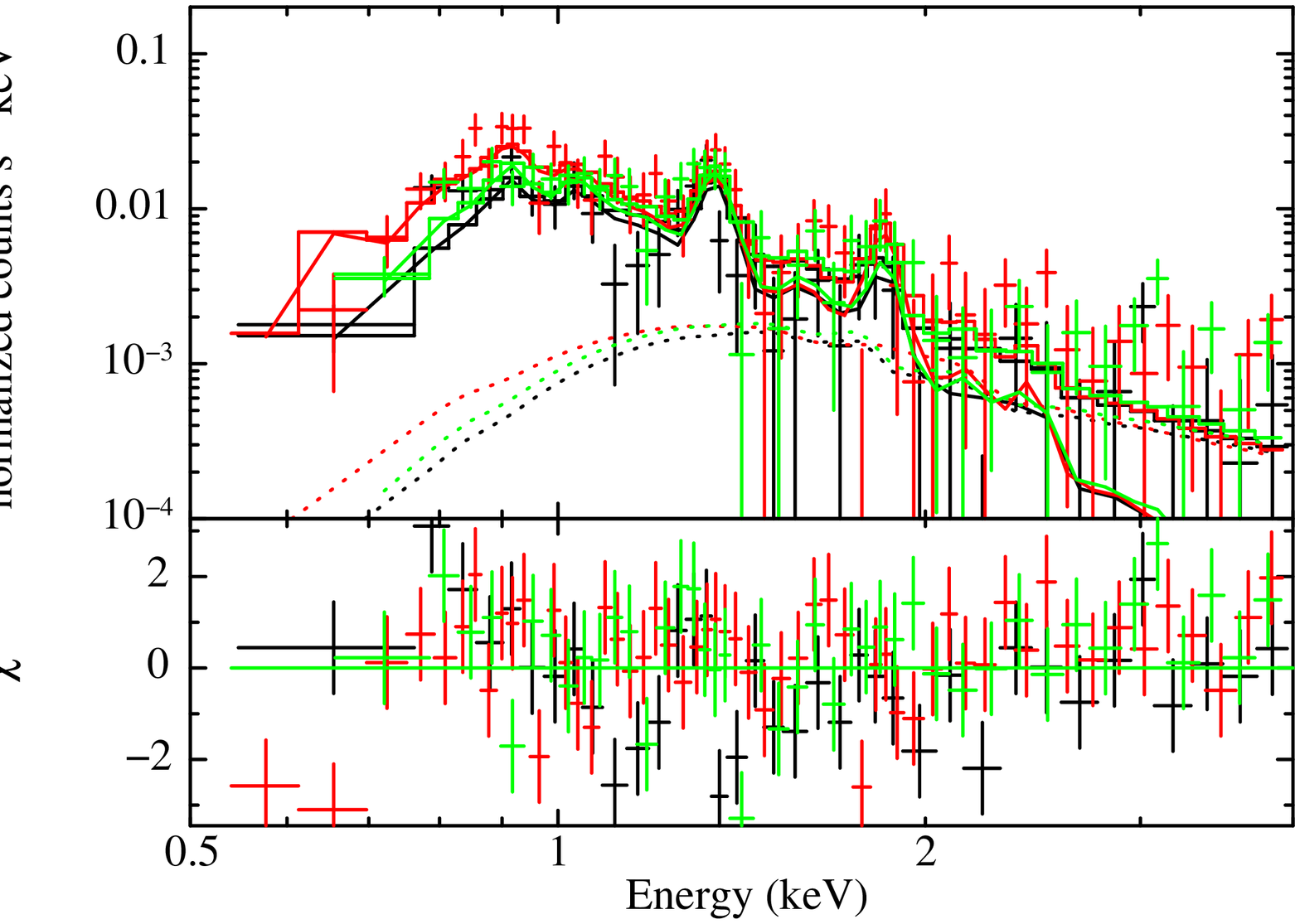}
\plotone{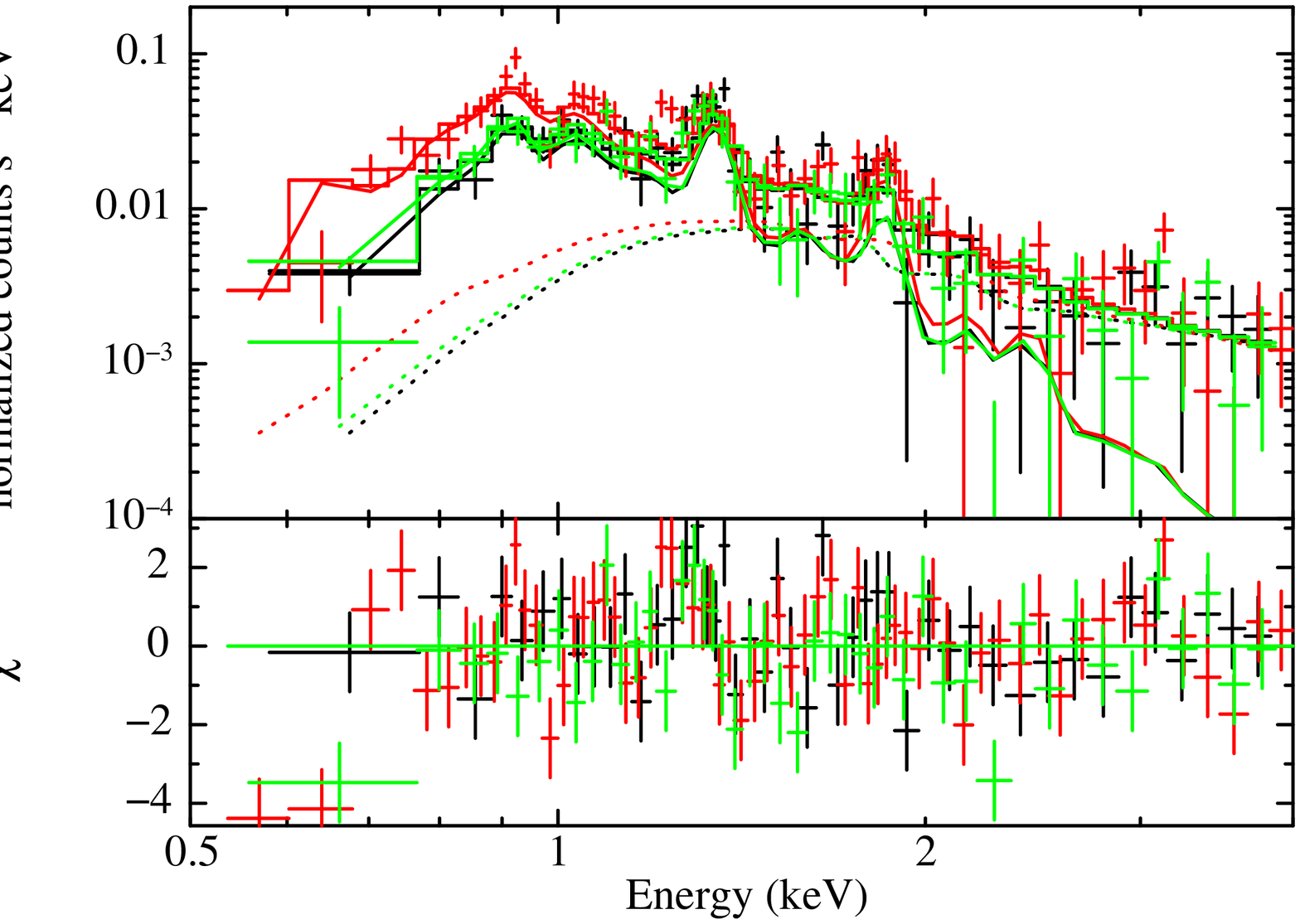}
\caption{%
Background-subtracted spectra of the north (left) and south (right)
region.
Black, red, and green crosses represent the XIS0, 1, and 3 specta,
respectively.
Solid and dotted lines represent {\tt vnei} and power-law component,
respectively.
}
\label{fig:diffuse_spectra_2}
\end{figure}

We also conducted a spatially resolved spectroscopy of the diffuse emission.
As shown in Figure~\ref{fig:xrayimage},
we divided the shell into a northern and southern region
% (see Figure~\ref{fig:xrayimage})
and extracted spectra for each region.
For the background subtraction, we used the simulated background for the total emission (as described above).
Due to the limited statistics,
we allowed the normalization of {\tt vnei} and power-law components to vary,
but the other parameters were fixed to the best-fit value of
the total emission.
The best-fit models and parameters are shown in 
Figure~\ref{fig:diffuse_spectra_2}
and Table~\ref{tab:diffuse_spectra_2}, respectively.
The fitting gave a similar reduced $\chi^2$/d.o.f to the total emission,
even with the smaller number of the free parameters.

\subsection{2XMM~J185114.3$-$000004}
\label{sec:2XMM}

\subsubsection{Spectral analysis}

\begin{figure}
\epsscale{0.5}
\plotone{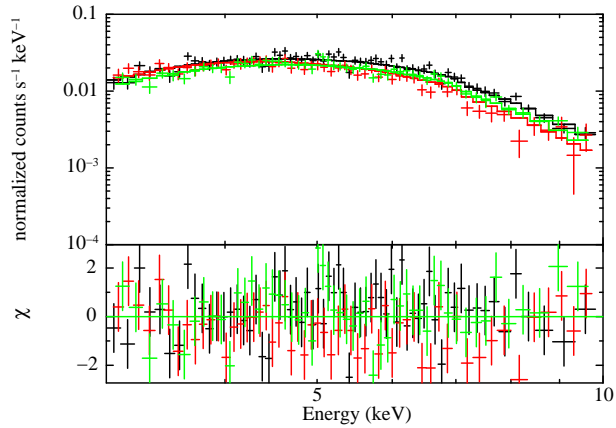}
\caption{%
Upper-panel:
Background-subtracted XIS spectra of 2XMM~J185114.3$-$000004
in the 3.0--10.0~keV energy band.
Lower panel: Residuals from the best-fit models.
In both panels, data in black, red, and green represent
the XIS0, 1, and 3 spectra, respectively.
}
\label{fig:2xmmspec}
\end{figure}

It is difficult to judge the emission on the 2XMM~J185114.3$-$000004 region
is point-like or not
due to the contamination of patchy thermal emission \citep{zhou2011}.
We assume it is a point source in the following analysis,
and this assumption is supported by the time variability
as shown in \S\ref{sec:timing}.
We extracted the source photons from a 3~arcmin-radius region.
The background region is the source-free region
common to the diffuse emission analysis.
The background-subtracted spectrum is shown in Figure~\ref{fig:2xmmspec}.
Below $\sim$3~keV, we see line emission which we believe is associated with contamination from the thermal emission of the SNR.
This emission very difficult to subtract correctly due to its patchy distribution.
We thus used only the 3--10~keV band for the spectral analysis of the point source,
since the diffuse emission is significantly fainter than the point source
above 3~keV
(see Figure~\ref{fig:diffuse_spectra} and Figure~\ref{fig:2xmmspec}).
The spectrum extends up to 10~keV, is heavily absorbed,
and shows no line-like emission.
An absorbed power-law model was thus adopted as the spectral model.
The absorption model includes the cross sections of
\citet{balcinska-church1992}
with solar abundance \citet{anders1989}.
The power-law model yielded an acceptable fit with a hard photon index, $\Gamma$$\sim$1.6,
and a reduced $\chi^2$ of 191.5/175.
The best-fit parameters are listed
in the second column of Table~\ref{tab:spec_2xmm}.
We also checked whether there is any significant iron-line emission.
For that, we added a narrow Gaussian model with the center energy of 6.4~keV
in the fitting,
and found a tight upper-limit on the equivalent width of 15~eV.

\subsubsection{XIS Timing analysis}
\label{sec:timing}

For the timing analysis of this point source,
we also used the 3--8~keV band and extracted photons
from a 3-arcmin radius region.
%for the good statistics.
The left panel of Figure~\ref{fig:xislightcurve} shows
the light curve of 2XMM~J185114.3$-$000004, which
%The light curve is 
reveals a highly time-variable source that
decays slowly during the observation.
One can also see rapid flares with a timescale of a few hundred seconds.
The short flares look rather periodic
with a timescale of $\sim$7000~s,
but this variability is not significant in our dataset
with the null hypothesis of 20\%.
We exclude 
%that the flares are due to 
the thermal wobbling of the satellite as a possible source for the flaring
\citep{uchiyama2008},
since this tentative period is not the same as the Suzaku satellite's orbital period
($\sim$96~min).
We conclude that this is real time-variability of the source.

In order to examine the spectral changes during the flares,
we compared the count rates in each bin of the light-curve for the 3--5~keV and 5--8~keV bands
(i.e. using bands where the source photons are dominant compared with background photons).
The right panel shows the 3--8~keV count rate
versus the hardness ratio between these bands.
One can see that there is no strong trend,
implying that there is no significant spectral change
during the flares.

A coherent pulsation search was also performed
although the time resolution of XIS is only 8~s
\citep{koyama2007}.
No significant pulsations were found
in the scanned period range of 16--$10^4$~s
with the null hypothesis of 31\%.
% scanned.

\subsection{HXD analysis}

We also analyzed the HXD P-I-N type diode (PIN) dataset
to search for a signal from the source in the energy range above 10~keV.
After background (NXB+CXB) subtraction,
the remaining count rate in the 15--40~keV band
in each observation is 
$2.4\pm0.2\times 10^{-2}$ cnts~s$^{-1}$ for OBSID=507035010
and
$1.4\pm0.3\times 10^{-2}$ cnts~s$^{-1}$ for OBSID=507036010.
This is 9.0\% and 5.4\% of the NXB count rate
in this band,
respectively.
The systematic NXB uncertainty is 3--5\%
\citep{fukazawa2009},
thus we conclude that we detect significant emission
from the northern part.

The observing region is on the Galactic plane,
thus we have to carefully account for the Galactic Ridge X-ray Emission (GRXE).
\citet{yuasa2008} reports that
the 12--40~keV HXD PIN count rate of the Galactic center is
$\sim$0.5~count~s$^{-1}$.
\citet{yamauchi1993} shows that the surface brightness of GRXE is
a few percent of that in the GC region,
thus the expected count rate is $\sim 10^{-2}$ count~s$^{-1}$,
which is significantly lower than our detection.
Moreover, the detection level should be the same in the two observations
if the emission is from GRXE.
We thus conclude that the emission is not from the GRXE.

Figure~\ref{fig:pinlc} shows the PIN 15--20~keV light curve
after the subtraction of the NXB.
We can see the decay of the emission,
which is similar to the light curve of 2XMM~J185114.3$-$000004
below 10~keV (Figure~\ref{fig:xislightcurve}).
This result confirms that the emission is from 2XMM~J185114.3$-$000004.

We have extracted the PIN spectrum above 10~keV
as shown in the left panel of Figure~\ref{fig:pinspec}.
We fit it with an absorbed power-law model.
The absorption column density was fixed to 11.0$\times 10^{22}$~cm$^{-2}$,
under the assumption that the emission is from 
2XMM~J185114.3$-$000004.
The fit was acceptable with a reduced $\chi^2$ of 12.4/10.
The best-fit photon index and flux are
4.2 (1.7--6.7) and 3.3 (2.6--4.1)$\times 10^{-12}$~erg~cm$^{-2}$s$^{-1}$
in the 15--20~keV band, respectively.
The photon index is relatively soft, but consistent within the error range
of the XIS result on 
2XMM~J185114.3$-$000004,
which also supports our conclusion on the emission origin.
We thus carried out the combined spectral fitting of XIS and PIN
for 2XMM~J185114.3$-$000004
which is shown in the right panel of Figure~\ref{fig:pinspec}.
We fit the spectra with an absorbed power-low model again,
and the fit was acceptable with a reduced $\chi^2$ of 208.6/183.
The best-fit models and parameters are shown in
the right panel of Figure~\ref{fig:pinspec} and
the third column in Table~\ref{tab:spec_2xmm}.
We note that we also attempted a absorbed broken-power-low model,
but this model did not improve the fit.

HXD has
a good timing capability with a 61~$\mu$s time resolution
and an accuracy of $1.9\times 10^{-9}$ss$^{-1}$ per day
\citep{terada2008}.
We thus search for any coherent pulsations of this source
using the PIN dataset.
We used the 15--20~keV range to maximize the signal-to-noise ratio.
After barycentric correction, we used the {\tt powspec} command
in the XRONOS package to search the coherent pulsation.
However, we could not detect any coherent signal
in the period of 61~$\mu$s to 512~s.
We also tried the timing analysis with the XIS (3--8~keV)
 and PIN (15--20~keV) combined,
and found no significant signal.

\section{Discussion}
\label{sec:discuss}

\subsection{Diffuse emission from the SNR}

We have detected the X-ray shell structure from G32.8$-$0.1
for the first time.
The best-fit absorption column of $5.9~(4.8-7.1)\times 10^{21}$~cm$^{-2}$
is marginally smaller than
the absorption column in this direction, 1.5--1.9$\times 10^{22}$~cm$^{-2}$.
Assuming an average density of
0.5--1~cm$^{-3}$,
the expected distance is in the 1.5--4.6~kpc range.
This is roughly consistent with the distance estimated from the CO association.
We thus use hereafter 4.8~kpc for the distance to the remnant.
The size of the X-ray shell is 14$^{\prime}$$\times$22$^{\prime}$,
which coincides with the radio shell.
The physical size and total luminosity are $20\times 31$~pc
and $\sim 2\times 10^{34}$~erg~s$^{-1}$ (0.5--10~keV),
assuming the kinematic distance of 4.8~kpc.

Next, we discuss the lower-temperature and hard-tail components
separately.

\subsubsection{Lower-temperature component}

In order to estimate the density and shock age of the low-temperature plasma
in this SNR,
we assumed that the shape of the emitting plasma is
an ellipsoid shell with dimensions of 7, 7, 10~arcmin,
with a width of 0.7~arcmin,
which correspond to 10, 10, and 15~pc with a width of 1~pc (at the assumed distance of 4.8~kpc).
%from the X-ray image and the distance assumption of 4.8~kpc.
The width of 1/12th the SNR radius is a rough estimation from the Sedov solution.
The total volume is then $4.5\times 10^{58}D_{\rm 4.8 kpc}^3$~cm$^{3}$,
where $D_{\rm 4.8 kpc}$ refers to the distance in the unit of 4.8~kpc.
Assuming a uniform density over the region,
we can derive the average electron density ($n_e$) from the emission measure as
\begin{equation}
n_e = 8.6~(6.1-13.6)\times 10^{-2} f^{-1/2}D_{\rm 4.8~kpc}^{-1/2}\ \ ({\rm cm}^{-3})
\end{equation}
where $f$ is the volume filling factor and we assumed $n_e = 1.2 n_{\rm H}$.
We used the {\tt vnei}+power-law model here (see Table~3)
%and assuming 
(the two-temperature {\tt nei} model will not change our parameters estimate within error).
The ambient density $n_0$ is 1/4 of $n_e$,
thus
\begin{equation}
n_0 = 2.2~(1.5-3.4)\times 10^{-2} f^{-1/2}D_{\rm 4.8~kpc}^{-1/2}\ \ ({\rm cm}^{-3})
\end{equation}
This is a rather low density
when compared with interstellar medium and other SNRs
even with the smaller distance estimate inferred from the absorption.
This is suggestive of expansion of G32.8$-$0.1 into its progenitor's wind bubble.
This is a similar situation to RCW~86,
which expands into the wind bubble with the density of
$10^{-3}$--$10^{-2}$~cm$^{-3}$ \citep{broersen2014},
although RCW~86 is much younger than G32.8$-$0.1 \citep{vink2006}.
The anti-correlation with the molecular clouds
and existence of OH masers further support this scenario.
Together with the $n_e$ estimate and the ionization time-scale
derived from the spectral analysis,
we can estimate the shock age of this plasma.
Adopting $n_e$ of $8.6\times 10^{-2}f^{-1/2}D_{\rm 4.8~kpc}^{-1/2}$,
the resultant shock age ($t$) becomes
\begin{equation}
2.2~(1.3-6.3) \times 10^4 f^{1/2}D_{\rm 4.8~kpc}^{1/2}\ \ (yrs)\ \ ,
\end{equation}
which suggests that G32.8$-$0.1 is a middle-aged SNR.
Its low luminosity further supports this conclusion
\citep{long1983}.

In the spatially resolved spectroscopy,
we derived the emission measure of thermal emission
in the northern and southern regions (Table~\ref{tab:diffuse_spectra_2}).
The surface brightness is similar in each region
since the area scale of the northern region is around 40\% that of the southern region.
Together with the same temperature in these regions,
we conclude that there is no substantial difference in their density.
%we have no clue of the environment difference in these regions.
Note that our observations did not fully cover the eastern part of the remnant,
where the molecular cloud dominates.
A more complete mapping of the remnant will reveal a clearer correlation
between the morphology of the molecular cloud and that of the SNR.

\subsubsection{The hard-tail component}

We have detected a significant hard-tail 
from the shell region of G32.8$-$0.1.
In the following, we discuss the origin of this emission.

The first possibility we consider is the contamination of hard point sources
in this region,
since the Suzaku angular resolution is not so excellent,
around 2~arcmin in half power diameter.
We have checked the {\it ROSAT} PSPC image in this region
and did not find any bright point source in the field.
Furthermore, the power-law component is significant in
both the southern and northern SNR regions,
which would be difficult to explain with point source contribution.
We thus conclude that this emission is not from a point source,
although we can not rule out the existence of obscured hard sources.

The second possibility is that
it is truly diffuse, but caused (at least partly) by the mis-estimation of the cosmic background level.
As shown in Figure~\ref{fig:background-spec},
the main background emission above 2~keV is 
the high-temperature component of the GRXE.
The fitting with the two-temperature {\tt vnei} model
shows that the temperature is consistent with the high-temperature component
of the GRXE.
We thus checked whether the power-law component can be 
explained by the GRXE high-temperature component,
and found that we need 27\% increase in the flux of the GRXE's high-temperature component
to reproduce this power-law component.
It is slightly a higher fluctuation than
the dispersion of GRXE (Uchiyama, H., 2014, private communication,
see also \cite{uchiyama2013}),
but we cannot dismiss this possibility.
We need more statistics in the hard X-ray band to distinguish this scenario.

An interesting possibility is a PWN origin.
The photon index and luminosity are typical for 
middle-aged PWNe \citep{kargaltsev2008,mattana2009}.
A PWN origin for the H.E.S.S. emission from the SNR region has been also suggested (Kosack et al. 2011).
However it is hard to explain both the northern and southern regions
having a power-law component;
on the other hand, it is possible that at least part of the emission has a PWN origin.
High-resolution X-ray observations with $Chandra$ will be needed to resolve the emission
and find any putative PWN or contaminating hard point sources.
%with better spatial resolution
%will allow us to point out where is the central pulsar,
%if the scenario is in case.

The more interesting possibility is that
the hard-tail emission is truly diffuse emission originating from the SNR.
In the scenario where the hard-tail is of thermal nature,
the high $kT$ and very small ionization timescale ($n_et$, see Table~3),
imply that the plasma was recently heated and has a low density.
Similar high $kT$ and small $n_et$ plasma
was found in RCW~86 \citep{yamaguchi2011};
it is the ejecta heated by the reverse shock very recently.
This may be the case of G32.8$-$0.1,
although it is much older than RCW~86
assuming the latter is associated with SN~185 \citep{vink2006}.
This scenario fits the multi-wavelength observations, and is consistent with
the picture that the SNR is expanding into a cavity and has hit the molecular cloud recently.
%expected from the OH maser.
%

Next we consider the non-thermal interpretation of the hard tail as originating from the SNR shell.
Shocks of SNRs accelerate particles up to 10$^{12}$~eV,
and accelerated electrons emit synchrotorn X-rays
\citep{koyama1995}.
The power-law component in G32.8$-$0.1 can be the synchrotron emission
from accelerated electrons.
This scenario is supported by the TeV detection.% and candidate GeV emission.
The luminosity of the hard component,
9.7$\times 10^{32}D_{\rm 4.8~kpc}^2$ in the 2--10~keV band,
is however rather small for synchrotron X-rays from young SNRs
\citep[$10^{32}-10^{36}$~ergs~s$^{-1}$ in the 2--10~keV band;][]{nakamura2012},
which further supports the middle-aged scenario.
An explosion in a cavity would keep the shock velocity high for longer
which would then lead to a high maximum energy of electrons
\citep{aharonian1999,yamazaki2006,zirakashvili2007}.
The average shock velocity inferred from the shock age and the SNR radius
is 5--7$\times 10^7$~cm~s$^{-1}$,
which is also consistent with a middle-aged SNR
and slower than typical value to emit synchrotron X-rays
\citep[noting that recent estimates for the shock speeds
in RCW~86 give values ranging from 700~km~s$^{-1}$ to 2200~km~s$^{-1}$,][]
{helder2013}.
%supporting our scenario.}
%Such a scenario would explain acceleration of particles even from such an aged system.
Since synchrotron X-rays from shocked plasma are expected to have
very thin filament-like structures
\citep{vink2003,bamba2003,bamba2005},
%due to the amplified and turbulent magnetic field.
X-ray observations with excellent spatial resolution, such as with {\it Chandra},  are needed to confirm this scenario
and would further allow for an estimate of the magnetic field.

\subsubsection{Comparison with other SNRs with gamma-rays}

Several SNRs have been detected in the GeV range by {\it Fermi} (see Funk 2011 for a review).
%\citep[][for the review]{funk2011}.
Their soft GeV emission may suggest the escape of high energy particles
\citep{ohira2011,ellison2011,telezhinsky2012,nava2013}.
Many of these sources are associated with molecular clouds,
and  particles accelerated at the shock emit gamma-rays
via pion decay with the dense material in the molecular clouds.
Our target G32.8$-$0.1 is interacting with a molecular cloud
(Figure~\ref{fig:multi-band}),
which is further supported by the existence of OH masers \citep{green1997}.
However, the association between the molecular cloud and GeV gamma-ray emission
is unclear;
the peak of the molecular cloud is in the eastern and northern region
whereas the gamma-ray peak is on the western side
without any overlap with molecular cloud and X-ray emission.
(Figure~\ref{fig:multi-band}).
%In order to examine the association of GeV emission for this source,
%more observations in the GeV range will be needed,
%although we should keep in mind that
%\citet{auchettl2014} suggest no convincing detection in the GeV band.
The VHE TeV gamma-ray emission, on the other hand,
coincides with the molecular cloud peak
in the eastern part of the SNR,
and is therefore likely associated with the SNR.

Here, we introduce two SNRs with gamma-rays,
W28 and RCW~86 for a comparison to G32.8$-$0.1.
W28 is a middle-aged SNR with GeV and VHE gamma-rays
\citep{abdo2009w28,aharonian2008}
detected on the north-eastern X-ray shell \citep{nakamura2014,zhou2014}.
\citet{nakamura2014} suggest that
the GeV/VHE gamma-ray emission is from high-energy particles
that recently escaped from the shock
and are interacting with a molecular cloud, giving also rise to 
OH masers \citep{frail1994,claussen1997,hoffman2005}.
The X-ray emitting plasma in the shell region
is in ionization equilibrium \citep{nakamura2014},
in contrast to the central emission \citep{sawada2012}.
The VHE gamma-rays from G32.8$-$0.1 may have a similar origin to W28.
\citet{zhou2014} suggest the presence of a  hard X-ray tail in the W28 shell,
which may be related to the hard tail in G32.8$-$0.1.
Another example, RCW~86,
is a bright shell-like SNR in the radio and X-ray bands
\citep{whiteoak1996,bamba2000}
expanding in a cavity \citep{broersen2014}.
This SNR has non-thermal emission in X-rays, GeV, and VHE gamma-rays
\citep{bamba2000,yuan2014,aharonian2009},
which is similar to our case.
On the other hand, we have no information on the molecular cloud interaction.
More data for such a sample are needed
to understand the multi-wavelength and intrinsic properties of these sources.

Young SNRs with VHE gamma-rays often have no significant thermal X-rays
\citep{koyama1997,slane2001,bamba2012},
suggesting a low-density environment.
Such a low density would keep a fast-moving shock speed for a relatively long time
and would accelerate particles more efficiently.
We suggest a similarly low density here for G32.8$-$0.1.

\subsection{2XMM~J185114.3$-$000004}

\citet{barthelmy2012} reported the Swift Burst Alert Telescope
detection of the outburst of 
2XMM~J185114.3$-$000004
on 2012 Jun 17 15:46:55 UT.
The mean count-rate in the
promptly-available XRT data is 0.56 count~s$^{-1}$,
whereas the catalogued count-rate of this source is equivalent to
approximately 0.013 XRT count~s$^{-1}$.
No optical counterpart has been found with the UVOT observation
onboard {\it Swift}.
We also searched for the infrared counterpart
in the 2MASS point source catalog \citep{skrutskie2006}.
The nearest 2MASS source is 2MASS 18511447$-$0000036,
but the separation between these sources is 2.6~arcsec,
which is larger than the position error of the 3XMM source (0.63~arcsec)
and the 2MASS source (0.08~arcsec).
We conclude that the source has no infrared counterpart.

\begin{figure}
\epsscale{0.45}
\plotone{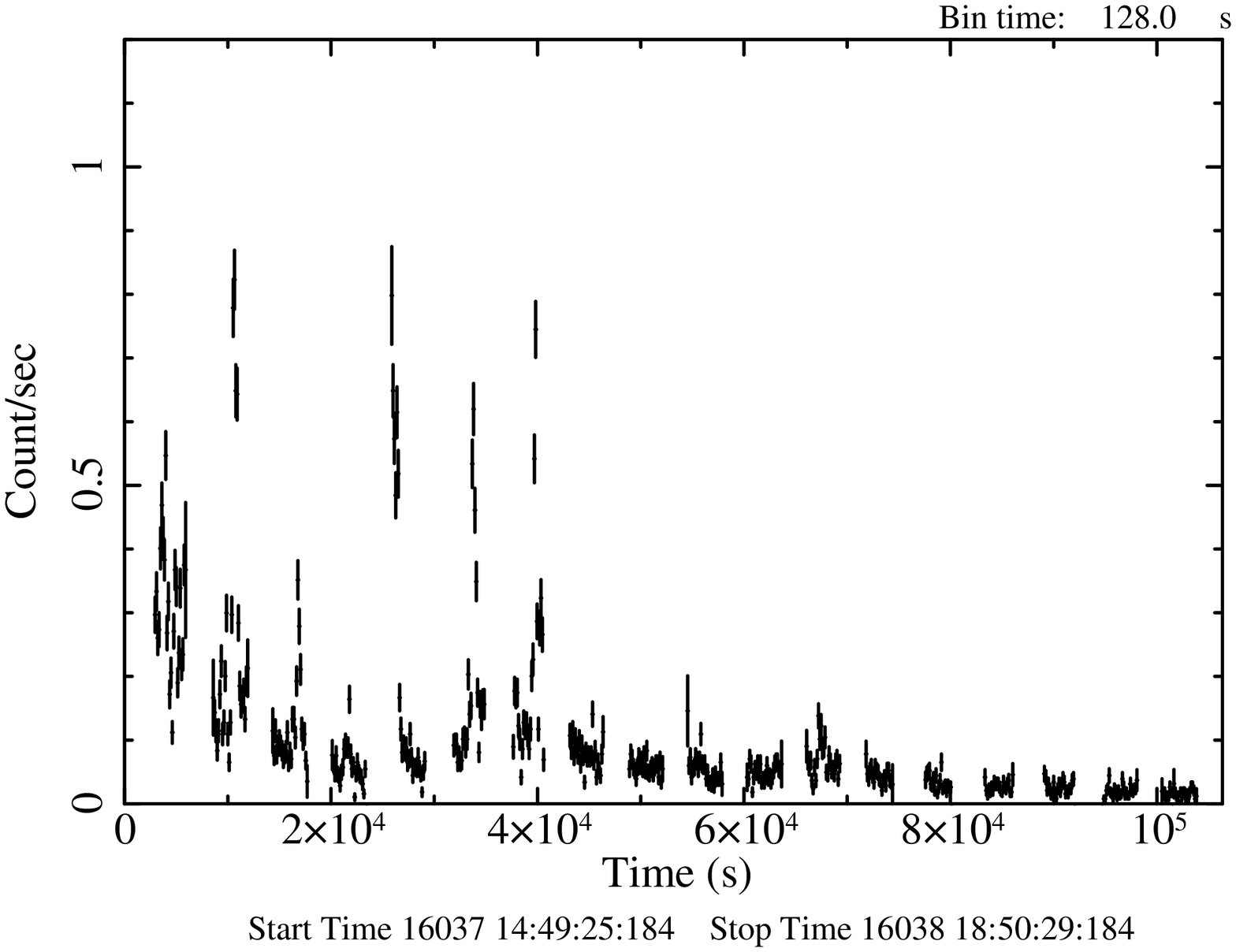}
\plotone{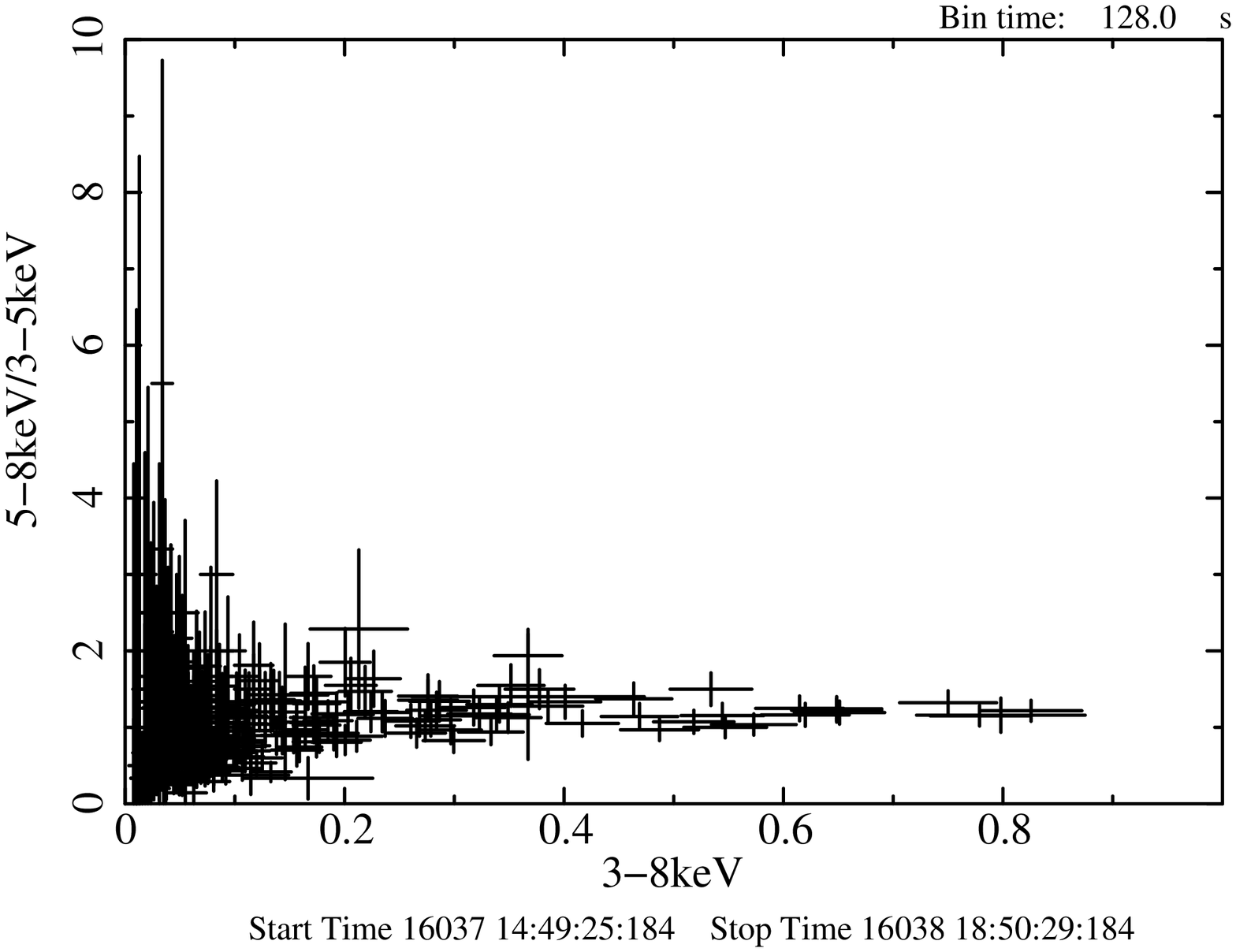}
\caption{%
Left: The 3.0--8.0~keV light curve of 2XMM~J185114.3$-$000004.
Right: 3--8~keV count rate vs. hardness ratio between 5--8~keV and
3--5~keV.
Bin time is 128~s in each panel.
}
\label{fig:xislightcurve}
\end{figure}

\begin{figure}
\epsscale{0.5}
\plotone{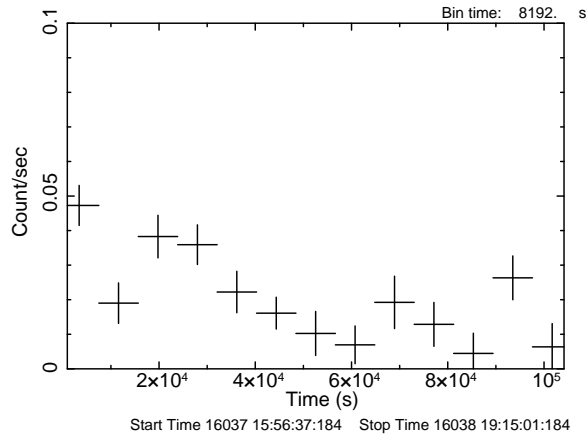}
\caption{%
The PIN 15--20~keV light curve in the OBSID=507035010 observation.
Bin time is 8192~s.
}
\label{fig:pinlc}
\end{figure}

\begin{figure}
\epsscale{0.45}
\plotone{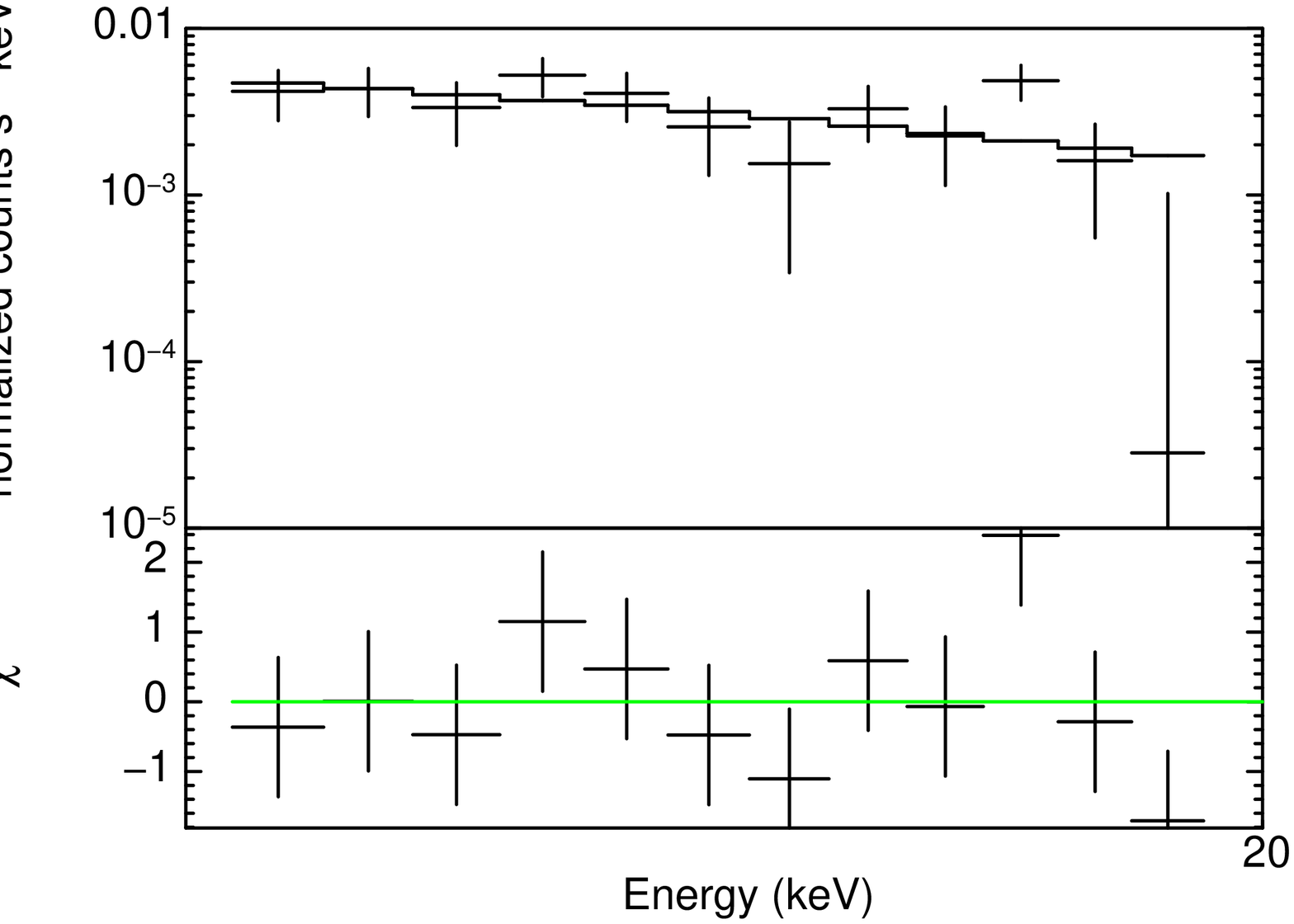}
\plotone{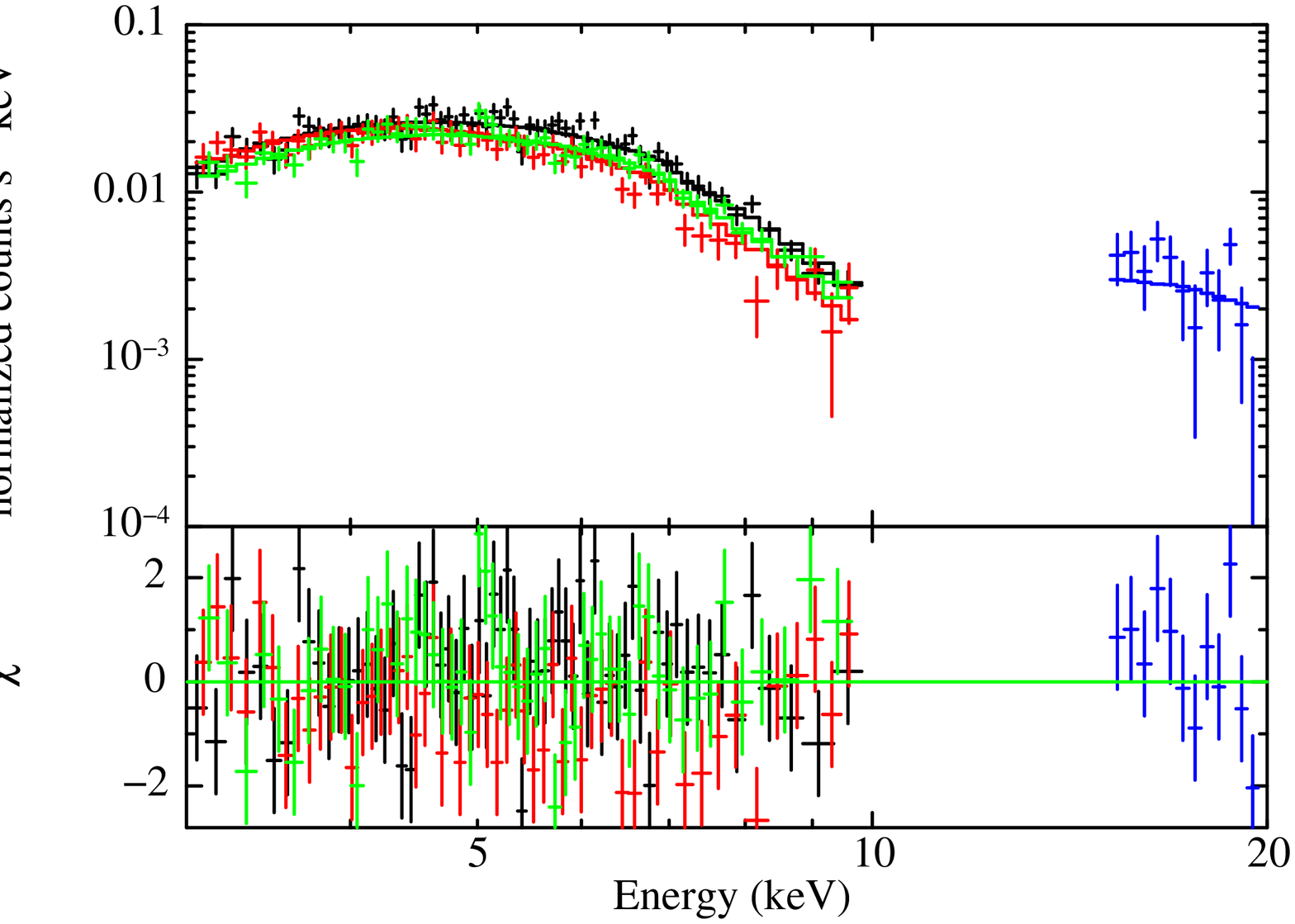}
\caption{%
Left panel: The background subtracted PIN spectrum in the 15--20~keV band.
Right panel: The wide-band spectra of 2XMM~J185114.3$-$000004.
Data in black, red, green, and blue represent
XIS0, 1, 3, and PIN spectrum.
In both panels,
the lower panels show the residuals from the best-fit model.
}
\label{fig:pinspec}
\end{figure}

With the assumption that the spectral shape is same
in the {\it Swift} observation and our observation,
we derived the 2--10~keV flux in the {\it Swift} observation
to be $7.7\times 10^{-11}$~erg~cm$^{-2}$s$^{-1}$
using {\tt webpimms}.
This is 1.5--1.6~count~s$^{-1}$ in {\it Suzaku} XIS 3--10~keV band.
These values are about one order of magnitude larger than
the average flux in the {\it Suzaku} observation
(table~\ref{tab:spec_2xmm}),
and twice the peak flux (Figure~\ref{fig:xislightcurve}).

The absorption column of 2XMM~J185114.3$-$000004 is
significantly larger than that of the SNR,
implying that this point source is unrelated to the SNR.
Assuming a distance of 10~kpc,
the intrinsic average luminosity is $1.1\times 10^{35}$~erg~s$^{-1}$
in the 2--10~keV band.
The absorption column is much higher than 
the total Galactic H~I column density toward the source
\citep[1.5--1.8$\times 10^{22}$~cm$^{-2}$;][]{kalberla2005},
which implies the source has a local absorption matter.

Recently,
a new class of Galactic transients has been emerging
with fast and bright flaring X-ray activity,
which is referred to as the supergiant fast X-ray transients
\citep[SFXT;][for a review]{sidoli2011}.
These objects are believed to be high-mass X-ray binaries
with (a few-hour long) hard X-ray spectra,
and short and bright flares.
The fast and large time variability, and the hard spectra of 
2XMM~J185114.3$-$000004, suggest that
this source is one of the SFXTs.
A rather large luminosity and no strong spectral change
further support this scenario \citep{bamba2001},
although \citet{kawabata2012} reported the change of hardness
in the case of AX~J1841.0$-$0536.
Some SFXTs show coherent pulsations from 4.7~s \citep{bamba2001}
to 1276~s \citep{walter2006},
%to 228~s \citep{lutovinov2005},
but our analysis did not show any evidence for pulsations.
The short timescale flares within a few hundreds second is 
common in other SFXTs \citep{walter2006}.
The origin of the ~7000~s intervals of short flares is unknown.
This timescale is not the {\it Suzaku} orbital period
\citep[$\sim$96~min,][]{mitsuda2007},
so it is not an artificial feature.
It is longer than the coherent pulsations of other SFXTs,
or and shorter than their orbital periods
\citep[3.3--165~days;][]{jain2009,walter2006}.

Another possibility is a flaring low mass X-ray binaries (LMXB).
Some LMXBs show rapid time variability and hard X-ray spectra
\citep[c.f., XSS~J12270$-$4859;][]{saitou2011,demartino2013}.
However, they show spectral hardening during flares,
whereas 2XMM~J185114.3$-$000004 didn't show any significant spectral change
during its flare.
Furthermore, the spectra of flaring LMXBs are not deeply absorbed,
which is also different from our source.
Recently, \citet{reig2012} reported that HMXBs with slow pulsations
can be accreting magnetars,
and our source can be a similar source.
However, such sources show only interstellar absorption,
which is not the case here.
A gamma-ray binary can be a candidate for our source
given their high variabilities and hard spectra.
However, these objects show spectral hardening when they become brighter
\citep[for example]{kishishita2009},
which is again not in the case of 2XMM~J185114.3$-$000004.
In summary, follow up high-resolution and deep observations are needed
to improve our understanding of this intriguing source.

\acknowledgments

We would like to thank the anonymous referee for the fruitful comments.
We thank T.~Sakamoto for comments on {\it Swift} data.
We also thank M. Ishida for comments on the time variability
of the background due to the thermal wobbling of the satellite.
This work was supported in part by
Grant-in-Aid for Scientific Research
of the Japanese Ministry of Education, Culture, Sports, Science
and Technology (MEXT) of Japan, No.~22684012 and 15K05107 (A.~B.).
S.S.H. acknowledges support from the Canadian Space Agency and
from the Natural Sciences and Engineering Research Council
of Canada (NSERC) through the Canada Research Chairs 
and Discovery Grants programs.
This research has made use of
NASA's Astrophysics Data System Bibliographic Services,
and the SIMBAD database,
operated at CDS, Strasbourg, France.

%{\it Facilities:} \facility{Suzaku}
\facility{Suzaku}

\onecolumn

\begin{deluxetable}{p{5pc}cccc}
\tabletypesize{\scriptsize}
\tablecaption{Observation Log
\label{tab:obslog}}
\tablewidth{0pt}
\tablehead{
\colhead{ObsID} & \colhead{Position} & \colhead{XIS Exposure} & \colhead{HXD Exposure}  \\
 & (J2000) & [ks] & [ks]
}
\startdata
507035010\dotfill & (282.8355, $-$0.0511) & 55.1 & 53.5 \\
507036010\dotfill & (282.8355, $-$0.0511) & 52.2 & 50.2 
\enddata
\end{deluxetable}

\begin{deluxetable}{p{8pc}c}
\tabletypesize{\scriptsize}
\tablecaption{Best-fit parameters of the spectral fitting of
the background spectra\tablenotemark{a}
\label{tab:background-spec}}
\tablewidth{0pt}
\tablehead{
\colhead{Parameters} & 
}
\startdata
CXB component & \\
\hspace{3mm}
$N_{\rm H}$ [$10^{22}$~cm$^{-2}$]\dotfill & 1.82 (fixed) \\
\hspace{3mm}
Photon index\dotfill & 1.4 (fixed) \\
\hspace{3mm}
Surface brightness\tablenotemark{b}\dotfill & 5.4$\times 10^{15}$ (fixed)\\
LHB component & \\
\hspace{3mm} 
$N_{\rm H}$ [$10^{22}$~cm$^{-2}$]\dotfill & 0.42 (0.14--0.69) \\
\hspace{3mm} 
$kT$ [keV]\dotfill & 0.07 (0.04--0.11) \\
\hspace{3mm} 
$E.M.$\tablenotemark{c}\dotfill & 0.16 (0.002--6.3)\\
GRXE component & \\
\hspace{3mm} 
$N_{\rm H}$ [$10^{22}$~cm$^{-2}$]\dotfill & 0.87 (0.84--0.94) \\
\hspace{3mm} 
$kT_{low}$ [keV]\dotfill & 0.59 (0.56--0.62)\\
\hspace{3mm} 
$E.M._{low}$\tablenotemark{c}\dotfill & 1.5 (1.2--1.8)\\
\hspace{3mm} 
$kT_{high}$ [keV]\dotfill & 6.3 (5.5--7.6)\\
\hspace{3mm} 
$E.M._{high}$\tablenotemark{c}\dotfill & 1.1 (1.0--1.2)\\
$\chi^2$/d.o.f.\dotfill & 389.0/236
\enddata
\tablenotetext{a}{Errors indicate single parameter 90\% confidence regions.}
\tablenotetext{b}{In the unit of erg~s$^{-1}$cm$^{-2}$arcmin$^{-2}$
in 2--10~keV band.}
\tablenotetext{c}{Emission measure in units of
$\frac{10^{-17}}{4\pi D^2}\int n_en_HdV$, where
$D$, $n_e$, and $n_H$ represent the distance, and electron and hydrogen densities, respectively.}
\end{deluxetable}

\begin{deluxetable}{p{8pc}ccc}
\tabletypesize{\scriptsize}
\tablecaption{
Best-fit parameters of spectral fitting of
the total diffuse emission spectra\tablenotemark{a}
\label{tab:total-spec}}
\tablewidth{0pt}
\tablehead{
\colhead{Parameters} & \colhead{vnei} & \colhead{vnei + vnei} & \colhead{vnei + power-law}
}
\startdata
$N_{\rm H}$ [$10^{22}$~cm$^{-2}$]\dotfill & 0.55 (0.48--0.61) & 0.74 (0.71--0.82) & 0.59 (0.48--0.71) \\
$kT_1$ [keV]\dotfill & 3.1 (2.3--5.0) & 0.63 (0.37--0.73) & 0.65 (0.44--0.97) \\
Ne\dotfill & 1 (fixed) & 1 (fixed) & 1.3 (1.1--1.5) \\
Mg\dotfill & 0.69 (0.55--0.94) & 1 (fixed) & 1 (fixed) \\
Si\dotfill & 0.41 (0.30--0.58) & 1 (fixed) & 1 (fixed) \\
Fe\dotfill & 0.22 (0.05--0.57) & 1 (fixed) & 0.28 (0.13--0.49) \\
nt [$10^{10}$cm$^{-3}$s]\dotfill & 1.3 (1.1--1.6) & 19 (11--59) & 6.0 (3.5--17.1) \\
$E.M.$\tablenotemark{b}\dotfill & 5.8 (4.2--7.5) & 10 (8.9--26) & 10 (5--25) \\
$kT_2$ [keV]\dotfill & --- & 3.4 (2.2--4.7) & --- \\
nt [$10^{9}$cm$^{-3}$s]\dotfill & --- & 2.7 (1.9--3.4) & --- \\
$E.M.$\tablenotemark{b}\dotfill & --- & 5.2 (4.0--9.2) & --- \\
$\Gamma$\dotfill & --- & --- & 2.3 (2.0--2.6) \\
$F_{pow}$ [$10^{-13}$~erg~cm$^{-2}$s$^{-1}$]\tablenotemark{c}\dotfill & --- & --- & 3.5 (2.8--4.2) \\
$\chi^2$/d.o.f.\dotfill & 232.0/122 & 208.3/122 & 223.1/121
\enddata
\tablenotetext{a}{Errors indicate single parameter 90\% confidence regions.}
\tablenotetext{b}{Emission measure in the unit of
$\frac{10^{-10}}{4\pi D^2}\int n_en_HdV$, where
$D$, $n_e$, and $n_H$ represent the distance, and electron and hydrogen densities.}
\tablenotetext{c}{In 2--10~keV band.}
\end{deluxetable}

\begin{deluxetable}{p{8pc}cc}
\tabletypesize{\scriptsize}
\tablecaption{Best-fit parameters of the spectral analysis of
northern and southern regions of the SNR shown in Fig.~1\tablenotemark{a}
\label{tab:diffuse_spectra_2}}
\tablewidth{0pt}
\tablehead{
\colhead{Parameters} & \colhead{north} & \colhead{south}
}
\startdata
{\tt vnei} $E.M.$\tablenotemark{b}\dotfill & 3.8 (3.5--4.1) & 6.5 (6.1--6.9) \\
$F_{pow}$ [$10^{-14}$~erg~cm$^{-2}$s$^{-1}$]\tablenotemark{c}\dotfill & 7.4 (4.6--10) & 27 (23--30) \\
$\chi^2$/d.o.f.\dotfill & 191.0/134 & 258.5/154
\enddata
\tablenotetext{a}{Errors indicate single parameter 90\% confidence regions.}
\tablenotetext{b}{Emission measure in units of
$\frac{10^{-10}}{4\pi D^2}\int n_en_HdV$, where
$D$, $n_e$, and $n_H$ represent the distance, and electron and hydrogen densities, respectively.}
\tablenotetext{c}{In the 2--10~keV band.}
\end{deluxetable}

\begin{deluxetable}{p{8pc}cc}
\tabletypesize{\scriptsize}
\tablecaption{Best-fit parameters of spectral fitting of
2XMM~J185114.3$-$000004\tablenotemark{a}.
\label{tab:spec_2xmm}}
\tablewidth{0pt}
\tablehead{
\colhead{Parameters} & \colhead{XIS only} & \colhead{XIS + PIN}
}
\startdata
$\Gamma$\dotfill & 1.61 (1.47--1.75) & 1.57 (1.44--1.69) \\
Flux [$10^{-12}$erg~cm$^{-2}$s$^{-1}$]\tablenotemark{b}\dotfill & 
9.8 (9.2--10.4) & 9.6 (9.0--10.2) \\
$N_{\rm H}$ [$10^{22}$H~cm$^{-2}$]\dotfill & 11.0 (9.7--12.2) & 10.7 (9.5--11.8) \\
$\chi^2$/d.o.f.\dotfill & 191.5/175 & 208.6/183
\enddata
\tablenotetext{a}{Errors indicate single parameter 90\% confidence regions.}
\tablenotetext{b}{In the 2--10~keV band.}
\end{deluxetable}

\end{document}